\documentclass[aps,prl,twocolumn]{revtex4}
\usepackage{amssymb}
\usepackage{graphicx}
\usepackage{amsmath}

\def\hom{\widehat{\omega}}

\usepackage{xcolor}
\usepackage{tikz}
\usepackage{stmaryrd}
\def\boris{}         
\def\borisb{}         
\def\kpara{\boldsymbol{k}}          %
\def\omegag{\omega_1}               %

\begin{document}

\title[]{A route to all frequency homogenization of periodic structures}

\author{Yan Liu, S\'ebastien Guenneau and Boris Gralak}
\address{CNRS, Ecole Centrale Marseille, Aix-Marseille Universit{\'e}, Institut Fresnel,
13397 Marseille Cedex 20, France}

\begin{abstract}
We start from a one-dimensional periodic multilayered stack in order to define a frequency power 
expansion of effective permittivity, permeability and bianisotropic parameters. It is shown from 
the first order that a simple dielectric multilayer can display a magnetoelectric coupling
effect and, from the second order, that artificial magnetism can be obtained with arbitrary
low contrast. However, this frequency power expansion is found to diverge at the first band gap
edge. Thus, an alternative set of effective parameters, made of the propagation index and the
surface impedance, is proposed. It is established that these effective parameters, as functions 
of the complex frequency, possess all the analytic properties required by the causality 
principle and passivity. Finally, we provide arguments to extend these results to the 
three-dimensional and frequency-dispersive case.
\end{abstract}

\label{firstpage}
\maketitle




There is currently a renewed interest in photonic crystals and
metamaterials \cite{Yablonovitch87,Ramakrishna08}, i.e. periodic structures exhibiting 
new phenomena such as 
negative refraction \cite{Veselago68,Pendry00}. The former composites are known to possess ranges
of frequencies (band gaps) for which no wave is allowed to propagate 
\cite{Yablonovitch87}, while the latter composites exhibit
a magnetic response associated with local RLC-circuit type \cite{Pendry99} or 
Mie \cite{OBrien02}, resonances. In this letter, we achieve such a magnetic activity 
with low-contrast dielectric layers.

Our proposal is based upon a definition of all frequency homogenization (AFH) for which it 
is necessary to propose parameters satisfying causality principle \cite{GT10,Wee2011}, i.e. 
analytic properties with respect to the complex frequency, and passivity.
This necessitates to go beyond the Fresnel inversion \cite{Decker09} which, for 
each fixed frequency and wavevector, directly translates 
reflectivity, transmission into effective permittivity, permeability and chirality. 
The extension of classical homogenization theory \cite{Bensoussan78} to high frequencies 
\cite{Craster10} is of pressing importance for physicists working in the 
field of photonic crystals and metamaterials in order to understand extra-ordinary properties 
such as artificial optical activity \cite{TretyakovPRB07} and magnetism \cite{Pendry99,OBrien02}. 
Applied mathematicians show a keen interest in this topic 
\cite{Felbacq05,Cherednichenko06, Craster10}, since periodic structures with small inductive and 
capacitive \cite{Pendry99} elements structured at sub-wavelength length scales (typically 
$\lambda/10~\mathrm{to}~\lambda/6$) \cite{Pendry99,OBrien02}, can clearly be regarded as almost 
homogeneous. 

The starting point of this letter is the transfer matrix $T(\omega,\kpara)$ associated with
a layer which represents the unit cell of a periodic multilayered stack. The frequency $\omega$ and
the wavevector $\kpara = (k_1,k_2)$ are the conserved
quantities of the system which is homogeneous with respect to time and space variables
$(x_1,x_2)$ [in this paper, we use an orthogonal set of coordinates
$(x_1,x_2,x_3)$ such that the layers are stacked in the $x_3$-direction, see Fig. \ref{figure1}].
The transfer matrix $T(z,\kpara)$ is an analytic function of the complex frequency 
$z$ ($\omega$ is thus the real part of $z$) and the wavevector $\kpara$. This analyticity 
property opens a route to the definition of effective homogeneous parameters for all 
frequency and wavevector spectrum. 

The matrix $T(z,\kpara)$ is first used to build directly usual
effective homogeneous parameters like the permittivity $\varepsilon_{\text{eff}}(z,\kpara)$, 
the permeability $\mu_{\text{eff}}(z,\kpara)$ corresponding to artificial magnetism, and a 
bianisotropy coefficient $\xi_{\text{eff}}(z,\kpara)$, hallmark of optical activity. The limit 
of this technique is then precisely defined: it is found that these effective 
parameters are appropriate only for frequencies ranging from the origin to the first band gap edge.

\begin{figure}[!h]
    \centering
\includegraphics{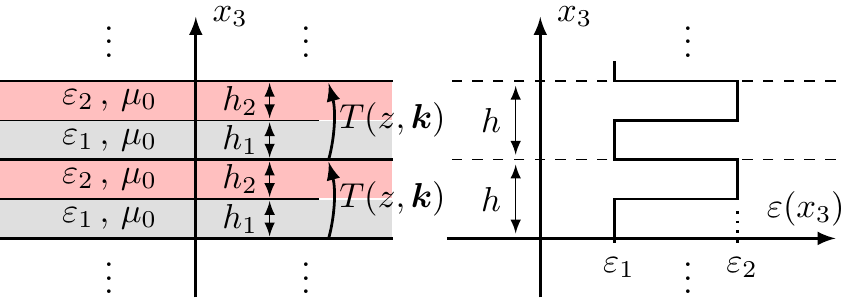}
\caption{Example of periodic multilayered stack of identical unit cells made of two dielectric 
homogeneous layers. The system is invariant under translations in the space directions $x_1$ 
and $x_2$ and the corresponding electric permittivity $\varepsilon(x_3)$ is a periodic function 
of variable $x_3$.}
\label{figure1}
\end{figure}

Next we turn to another set of effective parameters, the index
$n_{\text{eff}}(z,\kpara)$ and the impedance $\zeta_{\text{eff}}(z,\kpara)$. Using general 
arguments based on existence of Bloch modes and the local density of states, it is shown 
that these parameters are analytic functions in the upper half-plane of $z$, 
\boris{have a limit at infinite frequencies,} and that
imaginary part of $z n_{\text{eff}}(z,\kpara)$ and real part of $\zeta_{\text{eff}}(z,\kpara)$
are both positive. As the central result of this paper, it follows that both
effective index and impedance satisfy causality and passivity requirements.
This result leads us to the conclusion that a periodic multilayered stack can be replaced
by a frequency- and spatially-dispersive homogeneous effective medium for all the spectrum of
frequencies $\omega$ and wavevectors $\kpara$. The tool of choice for our one-dimensional model 
is the transfer matrix method, as it allows for analytical formulae, but we stress that ideas 
contained therein can be extended to frequency dispersive and three-dimensional periodic 
structures. 

We first consider a periodic multilayer with a unit cell made of two homogeneous layers $\mathcal{L}_1$
and $\mathcal{L}_2$ of thicknesses $h_1$ and $h_2$ ($h = h_1+h_2$), see Fig. \ref{figure1}. At the
oscillating frequency $\omega$, the electric and magnetic fields $\boldsymbol{E}$ and $\boldsymbol{H}$ 
are related to the electric and magnetic inductions $\boldsymbol{D}$ and $\boldsymbol{B}$ through the 
time-harmonic Maxwell's equations,
\begin{equation}
- i \omega \boldsymbol{D} (\boldsymbol{x}) = \boldsymbol{\nabla} \times \boldsymbol{H} (\boldsymbol{x}) \, , \quad
i \omega \boldsymbol{B} (\boldsymbol{x}) = \boldsymbol{\nabla} \times \boldsymbol{E} (\boldsymbol{x}) \, ,
\label{Maxwell}
\end{equation}
and the phenomenological constitutive relations for non-magnetic isotropic dielectric media:
\begin{equation}
\boldsymbol{D} (\boldsymbol{x}) = \varepsilon_m \boldsymbol{E} (\boldsymbol{x}) \, , \quad
\boldsymbol{B} (\boldsymbol{x}) = \mu_0 \boldsymbol{H} (\boldsymbol{x}) \, , 
\quad \boldsymbol{x} \, \in \, \mathcal{L}_m \, ,
\label{M0}
\end{equation}
where $\mu_0$ is the vacuum permeability, and $\varepsilon_m$ the permittivity in the
homogeneous layer $\mathcal{L}_m$, $m = 1,2$. Note that, at this stage, the frequency dependence 
of the permittivity is omitted. However, as it will be discussed later on, all the results of 
this letter remain valid when frequency dispersion is considered. In order to take advantage of 
invariance under translations in $(x_1,x_2)$, a Fourier decomposition [from $(x_1,x_2)$ to 
$\kpara = (k_1,k_2)$] is used to derive a first order differential equation from (\ref{Maxwell}):
\begin{equation}
\label{dFdx3}
\dfrac{\partial F}{\partial x_3} \, (\omega,\kpara,x_3)
= - i M_m(\omega, \kpara) \, F (\omega,\kpara,x_3) \, , 
\end{equation}
where $F$ is a column vector containing the tangential components of the Fourier-transformed electromagnetic 
field $(\,\widehat{\boldsymbol{\!E}},\,\widehat{\boldsymbol{\!H}})$, and $M_m(\omega,\kpara)$ is a
matrix independent of $x_3$ in each homogeneous layer $\mathcal{L}_m$. For all vector
$\boldsymbol{x}$, $x_\parallel$ is the component along the wavevector $\kpara = (k_1,k_2)$
and $x_\perp$ its component along $(-k_2,k_1)$. 
Then, omitting the $(\omega,\kpara)$-dependence, one has for $s$-polarization
\begin{equation}
F = \left[ \begin{array} {c} \widehat{E}_\perp \\ \omega \widehat{H}_\parallel \end{array} \right] \, , \quad
M_m = \left[ \begin{array} {cc}
0 & \mu_0 \\ \omega^2 \varepsilon_m - \kpara^2 / \mu_0 \quad& 0 \end{array} \right] \, .
\label{MatM}
\end{equation}
Since the matrices $M_m$ are $x_3$-independent, the solution of the equation (\ref{dFdx3}) in each layer
$\mathcal{L}_m$ is simply
\begin{equation}
\label{solution}
F(x_3+h_m) = \exp[ - i M_m h_m ] \, F (x_3) \, .
\end{equation}
The exponential above is well-defined as a power series of the matrix $M_m$, and defines the transfer
matrix in the medium $m$ through the distance $h_m$. Since this power series has infinite 
radius of convergence, the transfer matrix is analytic with respect to the three independent variables 
$\omega$, $k_1$ and $k_2$ describing the whole complex plane. From now on, the complex frequency will be
denoted by $z = \omega + i \eta$, where $\omega$ remains the real frequency and $\eta$ is the imaginary part.
The transfer matrix associated with the unit cell,
\begin{equation}
T(z,\kpara) = \exp[ - i M_2(z,\kpara) h_2 ]
\exp[ - i M_1(z,\kpara) h_1 ] \, ,
\label{MatT}
\end{equation}
is also analytic function in the whole complex plane of variables $z$, $k_1$ and $k_2$ (for arbitrary 
permittivity profile, analyticity is proved using a Dyson expansion \cite{RS2}).

The analyticity property opens the possibility to extract from the transfer matrix effective parameters which are
analytic functions of the complex frequency $z$ and valid over the whole frequency spectrum. First, the infinite
radius of convergence of the power series expansion of $T(z,\kpara)$ suggests to introduce
a notion of high order homogenization, 
which extends the usual homogenization (corresponding to the
limit $z \rightarrow 0$) by expanding the effective permittivity and permeability as power series of $z$.
To carry out the asymptotic analysis, we use the Baker-Campbell-Hausdorff formula (BCH, extension of the
Sophus Lie theorem, see \cite{Weiss62}):
\begin{equation}
\begin{array}{l}
\exp [A] \exp [B] = \exp[X] \, , \\[2mm]
X = A + B + \llbracket A,B \rrbracket + \llbracket  A-B , \llbracket A, B \rrbracket \rrbracket / 3  + \cdots \, ,
\end{array}
\label{BCH}
\end{equation}
where $A + B$ is defined as the zeroth order approximation (classical homogenization),
the commutator of $A$ and $B$ $\llbracket A,B \rrbracket = (AB-BA) / 2$ is the
first order approximation, $\llbracket  A-B , \llbracket A,B \rrbracket \rrbracket / 3$
is the second order approximation, and so forth. The BCH formula (\ref{BCH}) shows that 
the transfer matrix (\ref{MatT}) can be written as the one of a frequency- and 
spatially-dispersive homogeneous medium characterized by 
$X = - i M_{\text{eff}}(z,\kpara) h$. The resulting matrix $M_{\text{eff}}(z,\kpara)$ 
corresponds to the constitutive equations
\begin{equation}
\begin{array}{ll}
\!\!\widehat{\!\boldsymbol{D}}(\kpara,x_3) \!\!\!&= \varepsilon_{\text{eff}}(z,\kpara)
\, \, \widehat{\!\boldsymbol{E}}(\kpara,x_3) \!+\! i K_{\text{eff}}(z,\kpara) J
\, \widehat{\!\boldsymbol{H}}(\kpara,x_3) , \\[2mm]
\!\!\!\widehat{\boldsymbol{B}}(\kpara,x_3) \!\!\!&=\!\mu_{\text{eff}}(z,\kpara) \,\widehat{\!\boldsymbol{H}}(\kpara,x_3)
\!+\! i J K_{\text{eff}}(z,\kpara) \widehat{\boldsymbol{E}}(\kpara,x_3) .
\end{array}
\label{liu1}
\end{equation}
Here, matrix $J$ represents the 90 degrees rotation around the $x_3$-axis and, in the coordinate 
system $(x_\parallel,x_\perp,x_3)$, the effective permittivity and permeability are 
\begin{equation}
\varepsilon_{\rm eff} =
\text{diag}(\varepsilon_\parallel,\varepsilon_\perp,\varepsilon_3) , \quad
\mu_{\rm eff} = \text{diag}(\mu_\parallel,\mu_\perp,\mu_3) \; ,
\label{epmueff}
\end{equation}
while the bianisotropic parameter measuring the magnetoelectric coupling effect 
\cite{TretyakovPRB07,Guth2012} is given by
\begin{equation}
K_{\text{eff}}=\text{diag}(K_\parallel,K_\perp,0) .
\label{xieff}
\end{equation}
Although the system is isotropic in the plane $(x_1,x_2)$, the spatial dispersion induced by the 
$\kpara$-dependence introduces a difference between the coefficients $\varepsilon_\parallel$ and 
$\varepsilon_\perp$, $\mu_\parallel$ and $\mu_\perp$, and $K_\parallel$ and $K_\perp$ \cite{Landau}. 
At $\kpara = \boldsymbol{0}$, equalities $\varepsilon_\parallel(z,\boldsymbol{0}) = 
\varepsilon_\perp(z,\boldsymbol{0})$, and so forth\dots, are retrieved.


After some elementary algebra, collecting terms up to the second order approximation in (\ref{BCH})
with $A=- i M_2 h_2$ and $B=- i M_1 h_1$, we obtain the following homogenized coefficients \cite{supp} for
$\kpara = \boldsymbol{0}$:
\begin{equation}\begin{array}{l}
\varepsilon_\parallel (\hat z) = \varepsilon_1 f_1 + \varepsilon_2 f_2 + {\hat z}^2 f_1 f_2
(\varepsilon_1\!-\!\varepsilon_2)(\varepsilon_1 f_1\!-\!\varepsilon_2 f_2)/(6 \varepsilon_0) , \\[2mm]
\mu_\parallel(\hat z) = \mu_0 - {\hat z}^2 f_1 f_2 \mu_0 (\varepsilon_1\!-\!\varepsilon_2)
(f_1\!-\!f_2)/(6 \varepsilon_0) , \\[2mm]
K_\parallel(\hat z) = {\hat z}(\varepsilon_1\!-\!\varepsilon_2)f_1 f_2/(2\sqrt{\varepsilon_0/\mu_0}) ,
\end{array}
\label{parameff}
\end{equation}
with $\hat z = z h \sqrt{\varepsilon_0 \mu_0}$ the normalized complex frequency and $f_m = h_m/h$. It is stressed that 
magnetoelectric coupling comes from the odd order approximation in (\ref{BCH}), while artificial 
magnetism and high order corrections to permittivity emerge from even order approximation. These 
results are fully consistent with descriptions in terms of spatial dispersion  \cite{Landau,Agranovich2006} 
where, expanding the permittivity in power series of the wavevector, first order yields optical activity and 
second order magnetic response. This equivalence of these two descriptions 
(frequency and wavevector power series) is confirmed by considering a unit cell with 
a center of symmetry, for example a stack of three homogeneous layers (permittivity $\varepsilon_m$ and 
thickness $h_m$, $m=1,2,3$) with $\varepsilon_3 = \varepsilon_1$ and $h_3=h_1$. Extending 
(\ref{BCH}) to the case $\exp [A] \exp [B] \exp[A] = \exp[X]$, see \cite{supp}, it is found that 
$K_{\text{eff}}= 0$, and thus it is retrieved that both magnetoelectric coupling and optical activity 
vanish in a medium with a center of symmetry \cite{Landau}.

Expansion in power series of frequency provides a new explanation
for artificial magnetism and optical activity. Analytic expressions (\ref{parameff})
of effective parameters can be used to analyze artificial properties
(expressions for normal incidence up to order 6 are provided in \cite{supp}).
In particular, we show from (\ref{parameff}) that: artificial
magnetism, previously proposed with high contrast \cite{OBrien02,Felbacq05,Cherednichenko06}, can be obtained
with arbitrary low contrast; and magnetoelectric coupling, previously achieved in
$\Omega$-composites \cite{TretyakovPRB07}, can be present in simple one-dimensional
mulitlayers.

Nevertheless, this frequency expansion of effective parameters cannot be
used for frequencies higher than the frequency $\omegag$ at the first band gap edge \cite{supp}.
To show this limitation, we consider for the sake of simplicity a 3 layer unit cell with
a center of symmetry and purely real dielectric constants $\varepsilon_1 = \varepsilon_3$ 
and $\varepsilon_2$: effective parameters $\varepsilon_{\text{eff}}$ and $\mu_{\text{eff}}$ 
derived from (\ref{BCH}) are \boris{then} purely real and $K_{\text{eff}} = 0$. At the band gap 
edge $\omegag$, either $\varepsilon_{\text{eff}}$ or $\mu_{\text{eff}}$
must vanish to allow for sign shifting, while the other parameter must take an infinite
value: the power series expansion of $X = - i M_{\text{eff}} h$ in (\ref{BCH})
diverges at $\omegag$. Indeed, when calculating $X = \log \{ \exp[A] \exp[B] \}$,
it appears that the function $\log$ is not analytic when its argument ``vanishes'', which
introduces a branch point at $\omegag$: this branch point implies that the radius of
convergence of the power expansion is bounded by $\omegag$ (see Fig. \ref{figure2}).
However, it is possible to choose the branch cut of the complex logarithm
from the branch point $\omegag$ in the lower half complex plane (Fig. \ref{figure2}). 
This choice makes it possible to define effective
parameters which are analytic functions in the upper half plane of the complex frequency
$z$, as required by causality principle \cite{Landau}.


\begin{figure}
    \centering
\includegraphics{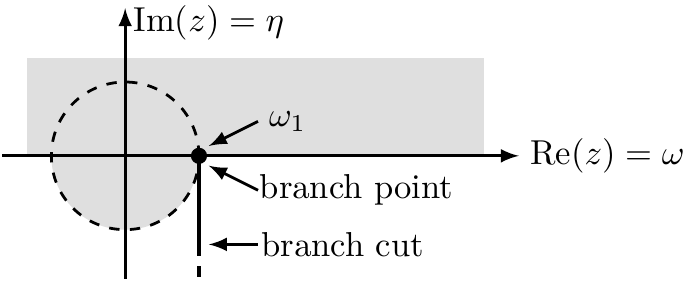}
\caption{Domains of analyticity (gray regions) of effective parameters
$\varepsilon_{\text{eff}}$, $\mu_{\text{eff}}$, $n_{\text{eff}}$ and $\zeta_{\text{eff}}$:
the circle \borisb{defines the disk} where the power expansion converges.}
\label{figure2}
\end{figure}

We consider a unit cell with a center of symmetry to keep things simple (no magnetoelectric 
coupling). The general expression of the transfer
matrix is then \cite{JMO2008} 
\begin{equation}
T(z,\kpara) = \left[ \begin{array} {cc} a(z,\kpara) & b(z,\kpara) \\
d(z,\kpara) & a(z,\kpara) \end{array} \right] \, , \quad a^2 - bd = 1 \, .
\label{MatTS}
\end{equation}
This matrix is compared with $T_{\text{eff}}(z,\kpara)$, the transfer matrix
corresponding to the constitutive equations (\ref{liu1}) with $K_{\text{eff}} = 0$
: omitting the $(z,\kpara)$-dependence, we have
\begin{equation}
T_{\text{eff}} = \left[ \begin{array} {lr} \cos[z n_{\text{eff}} h] &
\! \! \! -i (\zeta_{\text{eff}}/z) \sin[z n_{\text{eff}} h] \\
\!-i (z/\zeta_{\text{eff}}) \sin[z n_{\text{eff}} h] \! \! \! &
\cos[z n_{\text{eff}} h] \end{array} \right] ,
\label{MatTeff}
\end{equation}
where $z^2 n_{\text{eff}}^2 = z^2 \varepsilon_{\text{eff}} \mu_{\text{eff}} - \kpara^2$ and 
$\zeta_{\text{eff}} = \mu_{\text{eff}} / n_{\text{eff}}$.
Comparison of (\ref{MatTS}) and (\ref{MatTeff}) \borisb{gives}
\borisb{definitions} of propagation index $n_{\text{eff}}$ 
\borisb{(along $x_3$-axis)}
and surface impedance $\zeta_{\text{eff}}$:
\begin{equation}
\cos[z n_{\text{eff}}(z,\kpara) h] \!= a(z,\kpara) \, , \;
\zeta_{\text{eff}}(z,\kpara) = z\sqrt{ \dfrac{b(z,\kpara)}{d(z,\kpara)}} \, .
\label{nzetaeff}
\end{equation}
Here, notice that $z n_{\text{eff}}(z,\kpara)$ is just the Bloch wavevector. 
The main result of this letter is the following. \textit{In the upper half plane of $z$,
i.e. for Im$(z) >0$: i) imaginary part
of $z n_{\text{eff}}(z,\kpara)$ and real part of $\zeta_{\text{eff}}(z,\kpara)$ are 
positive; ii) $n_{\text{eff}}(z,\kpara)$
and $\zeta_{\text{eff}}(z,\kpara)$ are analytic functions of $z$; 
\boris{iii) $n_{\text{eff}}(z,\kpara)$
and $\zeta_{\text{eff}}(z,\kpara)$ have limits $n_\infty$ and $\zeta_\infty$ when 
$|z| \longrightarrow \infty$, where $n_\infty = \langle \sqrt{\varepsilon \mu_0} \, 
\rangle$ is the mean index.}}

To prove these claims, we first use the theorem stating that no Bloch mode exists for $z$
in the upper half plane \cite{TMC00}. As a consequence, the function $z n_{\text{eff}}(z,\kpara)$
cannot be purely real and its imaginary part Im$(z n_{\text{eff}})$ cannot vanish. This 
proves assertion i) for $n_{\text{eff}}$. Next, it is stressed that the coefficient $a(z,\kpara)$ 
is $z$-analytic in all the complex plane of $z$, and that the definition (\ref{nzetaeff}) 
can be written \begin{equation}
\exp[ i z n_{\text{eff}}(z,\kpara) h] = a(z,\kpara) + i \sqrt{1 - a^2(z,\kpara)} \, .
\label{neff}
\end{equation}
Since Im$(z n_{\text{eff}})$ cannot
vanish if Im$(z) >0$, we have $\cos[ z n_{\text{eff}}(z,\kpara) h]  = a(z,\kpara) \neq \pm 1$,
and thus the square root in (\ref{neff}) is $z$-analytic in the upper half plane.
The function on the left hand side of (\ref{neff}) is then analytic and, in addition,
cannot vanish. The complex logarithm can be applied to (\ref{neff})
without alteration of the analyticity property: this proves ii)
for $n_{\text{eff}}$. Next, combining the two equations $a^2-bd=1$ (\ref{MatTS}) and
$a\neq \pm 1$ 
for Im$(z)>0$, it is found that none of the two analytic functions $b(z,\kpara)$ and 
$d(z,\kpara)$ vanishes, and thus the ratio $b(z,\kpara)/d(z,\kpara) \neq 0$ is analytic. 
The square root in (\ref{nzetaeff}) preserves the analyticity property, which proves 
assertion ii) for $\zeta_{\text{eff}}$. The \boris{proof of i) for $\zeta_{\text{eff}}$} 
is based on the local density of states. The Green's function
of the multilayer is calculated \cite{GG07} with a point source located 
in the plane $x_3 = 0$ : the value of the electric field in the same plane is found to be
$ i \pi \zeta_{\text{eff}}$. Since the imaginary part of the Green's function
is positive (it corresponds to the local density of states), it follows that
$\zeta_{\text{eff}}$ has its real part Re$(\zeta_{\text{eff}})$ positive.
\boris{Finally, the proof of iii) is given in \cite{supp}.}

\boris{The main result tells us that artificial frequency dispersion, i.e. 
$z$-dependence (of effective parameters) generated by periodic spatial modulation, has the 
same properties as the natural frequency dispersion in usual media. Part i) ensures 
passivity, while parts ii) and iii) imply that the effective parameters fulfills 
causality principle \cite{Landau}. From ii) and iii), the Cauchy 
integral formula can be applied to the function $n_{\text{eff}}(z,\kpara) - n_\infty$ to 
obtain Kramers and Kronig relations and their generalization \cite{GT10}:}
\begin{equation}
n_{\text{eff}}(z,\kpara) = \boris{n_\infty} - \dfrac{1}{\pi} \displaystyle\int_{\mathbb{R}} d\nu  \, \,
\displaystyle\frac{\nu \, \mbox{Im}[n_{\text{eff}}(\nu,\kpara) \borisb{\, - \, n_\infty}]}{z^2 - \nu^2} \, .
\label{KKgen}
\end{equation}
\boris{An illustrative example in Fig. \ref{figure3} confirms that this generalization of the Kramers and 
Kronig relations is satisfied by $n_{\text{eff}}(z,\kpara)$, 
since the solid curve and plus markers of $\text{Re}(n_{\text{eff}})$ fit each other.
Also, Fig. 3 confirms that, at the infinite frequency limit, the effective index tends to the 
mean refractive index $\langle \sqrt{\varepsilon} \, \rangle$, as in usual media 
for very high frequencies of neutrons \cite{XrayPL}.}
Note that, contrary to frequency independent dieletric constants 
$\varepsilon_1 = \varepsilon_3 = 2 \varepsilon_0$ and $\varepsilon_2 = 12 \varepsilon_0$ of the 
multilayered stack, effective parameter $n_{\text{eff}}(z,\kpara)$ agrees with causality 
principle.

The main result remains valid when (natural) frequency dispersion is considered in the dielectric 
permittivity [e.g. dielectric constants $\varepsilon_m(z)$]. Indeed, permittivity is 
analytic in the upper half plane of $z$ and theorems on existence of Bloch modes and
imaginary part of Green's function can be applied in the most general cases
\cite{TMC00,GT10}. \boris{Note that, when natural frequency dispersion is considered, the limits 
$n_\infty$ and $\zeta_\infty$ take the values of index and impedance in vaccum, i.e.
$\sqrt{\varepsilon_0 \mu_0}$ and $\sqrt{\mu_0/\varepsilon_0}$.}

\boris{Finally, the possibility to extend our results to frequency dispersive 
three-dimensional periodic structures is discussed.} \borisb{Using the auxiliary field formalism
\cite{TMC00,GT10},} Maxwell's equations can be written as the unitary time evolution equation 
$[\partial F/\partial t] (t) = - i \mathsf{K} F(t)$,
where $\mathsf{K}$ is selfadjoint and time-independent (see also the frame of extension of 
dissipative operators \cite{Figotin2006}). Consequently the resolvent
$[z - \mathsf{K}]^{-1}$ is analytic if Im$(z)>0$, which prevents the existence of Bloch modes
in the upper half plane of $z$. And imaginary part of Green's
function is always positive for Im$(z)>0$ since the operator
\begin{equation}
-\dfrac{1}{2i}\left[ \dfrac{1}{z-\mathsf{K}} - \dfrac{1}{\overline{z} - \mathsf{K}} \right]
= \text{Im}(z) \, \dfrac{1}{z - \mathsf{K}} \, \dfrac{1}{\overline{z} - \mathsf{K}} \geq 0 \,
\end{equation}
is positive. Assuming that appropriate effective parameters can be defined 
(see promising notions of impedance in \cite{Lawrence2009,LalanneSmigaj,Simovski07}),
these two properties on Bloch modes and imaginary part of Green's function
can be used to prove causality principle and passivity.
It is however stressed that such a generalization remains a challenging task: particularly, 
special attention should be paid to situations where single mode Bloch approximation does not 
apply \cite{Simovski07}, and to analytic continuations of Bloch 
wavevector and impedance for complex $z$.

%
\begin{figure}
    \centering
    \resizebox{85mm}{!}{\includegraphics{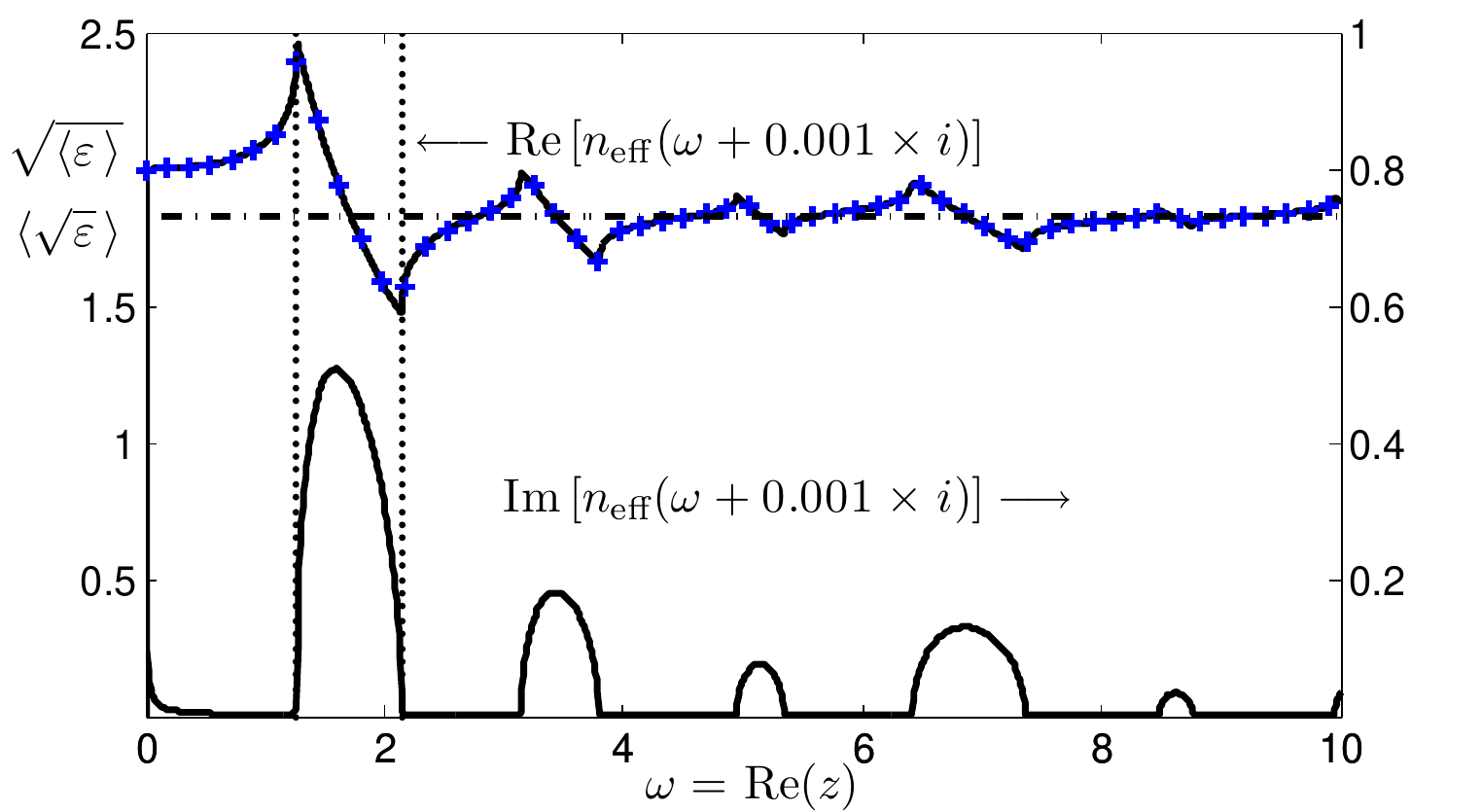}}
    \caption{Real and imaginary parts (solid lines) of effective refractive index $n_{\text{eff}}(z)$ \boris{for 
	$z = \omega + 0.001 \times i$, deduced from (\ref{neff}); Kramers and Kronig formula unveils $\rm Re(n_{\text{eff}})$ in plus markers; $\varepsilon_1 = \varepsilon_3 = 2 \varepsilon_0$,
    $\varepsilon_2 = 12 \varepsilon_0$, $f_1=f_3=0.4$, $f_2=0.2$, $\sqrt{\langle \varepsilon \, \rangle} = \sqrt{\varepsilon_1 f_1 + 
\varepsilon_2 f_2 + \varepsilon_3 f_3} = 2 \varepsilon_0$, $\langle \sqrt{\varepsilon} \, \rangle = \sqrt{\varepsilon_1} f_1 + 
\sqrt{\varepsilon_2} f_2 + \sqrt{\varepsilon_3} f_3 \approx 1.82 \varepsilon_0$.}}
\label{figure3}
\end{figure}

In conclusion, a periodic multilayered stack can be modelled at 
low frequencies by a homogeneous medium characterized by effective permittivity, permeability
and magnetoelectric coupling parameter. But these parameters, defined as power 
expansions, are not approriate for frequencies beyond the first band gap edge. Considering instead 
effective propagation index and surface impedance, it is shown that artificial frequency dispersion
has the same properties as natural dispersion in terms of passivity and causality: 
remarkably it follows that a periodic arrangement of frequency independent (and thus 
non-causal) dielectric materials makes artificial causality.
These results, based on general properties 
of existence of Bloch modes and sign of imaginary part of Green's function,
open a route for generalization of AFH to frequency 
dispersive and three-dimensional case.

\widetext{
\section{Supplemental Material}

\def\dfrac{\displaystyle\frac}
\def\ep{\varepsilon}
\def\om{\omega}
\def\hom{\widehat{\omega}}
\def\tr{\text{tr}}
\def\kpara{\boldsymbol{k}}

\subsection{Effective parameters for a one-dimensional periodic multilayer}

In this supplementary material we explain in more details our homogenization algorithm for the periodic multilayer as shown in Fig. \ref{figure1}. The effective medium is depicted in the right panel, with effective permittivity, permeability and bianisotropic parameters being rank-2 tensors.
\begin{figure}[!h]
    \centering
    \includegraphics[scale=0.6]{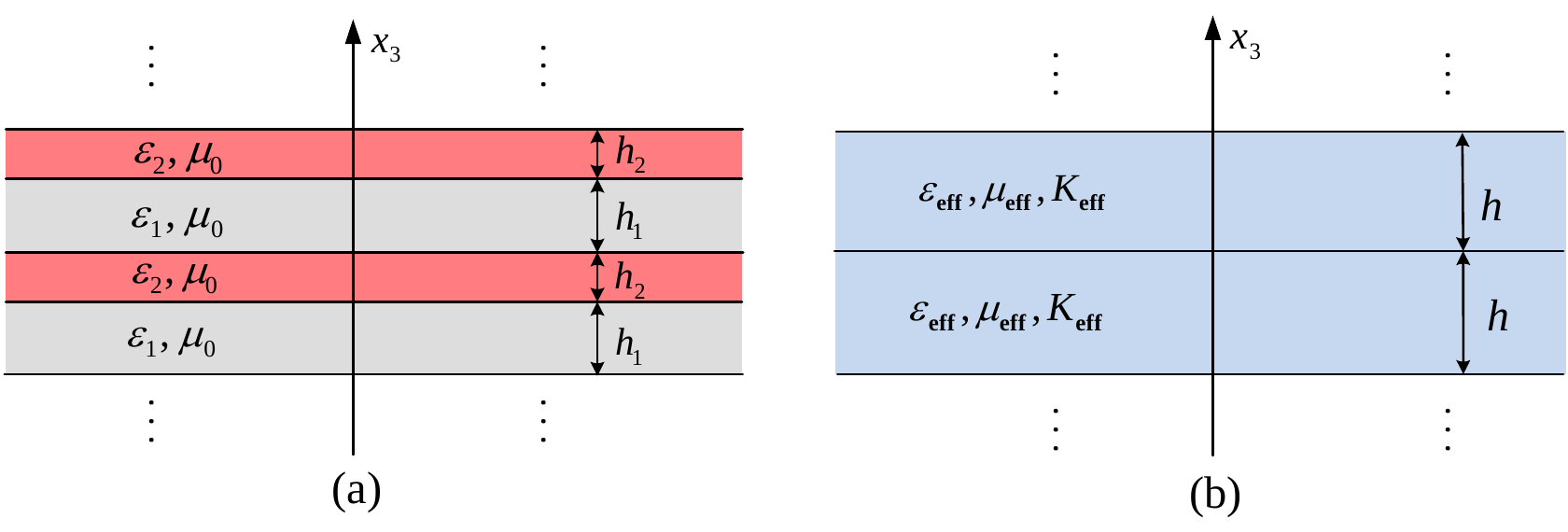}
    \caption{Schematic diagram of the periodic multilayer and corresponding effective medium.}
    \label{figure1}
\end{figure}
Let us start from the Maxwell's equations
\begin{equation}
- i \omega \boldsymbol{D} (\boldsymbol{x}) = \boldsymbol{\nabla} \times \boldsymbol{H} (\boldsymbol{x}) \, , \quad
i \omega \boldsymbol{B} (\boldsymbol{x}) = \boldsymbol{\nabla} \times \boldsymbol{E} (\boldsymbol{x}) \, ,
\label{Maxwell}
\end{equation}
and the constitutive relations for non-magnetic isotropic dielectrics
\begin{equation}
\boldsymbol{D} (\boldsymbol{x}) = \varepsilon_m \boldsymbol{E} (\boldsymbol{x}) \, , \quad
\boldsymbol{B} (\boldsymbol{x}) = \mu_0 \boldsymbol{H} (\boldsymbol{x}) \, , \quad \boldsymbol{x} \, \in \, \mathcal{L}_m \, ,
\label{cons}
\end{equation}
where $\mu_0$ is the vacuum permeability and $\varepsilon_m$ the permittivity in homogeneous layer $\mathcal{L}_m$.
Here we introduce the Fourier decomposition for both the electric and magnetic field in the form
\begin{equation}
\begin{array}{ll}
\widehat{\bf E}(k_1,k_2,x_3) \hspace{-3mm}& = \dfrac{1}{2 \pi} \displaystyle\int_{\mathbb{R}^2} {\bf E}(x_1,x_2,x_3) \exp\big[- i (k_1 x_1 + k_2 x_2)] \, dx_1 dx_2 \, , \\[4mm]
\widehat{\bf H}(k_1,k_2,x_3) \hspace{-3mm}& = \dfrac{1}{2 \pi} \displaystyle\int_{\mathbb{R}^2} {\bf H}(x_1,x_2,x_3) \exp\big[- i (k_1 x_1 + k_2 x_2)] \, dx_1 dx_2 \, .
\end{array}
\label{fourier}
\end{equation}
Applying the decomposition (\ref{fourier}) to (\ref{Maxwell}) and (\ref{cons}), we derive an ordinary differential equation involving $4\times 4$-matrices and a 4-components column vector\cite{lakhtakia92}:
\begin{equation}
\dfrac{\partial \textit{{\textbf F}}}{\partial x_3} \, (\omega,\kpara,x_3)
= - i M_m(\omega, \kpara) \,\textit{{\textbf F}} (\omega,\kpara,x_3) \, .
\label{diff}
\end{equation}
Here, the field-components $\widehat{E}_3$ and $\widehat{H}_3$ have been eliminated and the resulting column vector $\textit{{\textbf F}}$ contains the electric and magnetic field-components which are tangential to the interface of the multilayered stack, i.e. the components along $x_1$ and $x_2$.

In order to simplify the derivation, a coordinates rotation of the $(x_1,x_2)$-plane is used: The new coordinates are denoted by $x_\parallel$ and $x_\perp$ where the component $x_\parallel$ is along the wavevector $\kpara = (k_1,k_2)$ and $x_\perp$ is along $(-k_2,k_1)$. In other words, the new coordinates are related to the previous ones through the following transformation
\begin{equation}
  \left[\begin{array}{l}
  x_\parallel \\
  x_\perp
  \end{array}
  \right]= \dfrac{1}{\sqrt{k_1^2+k_2^2}}\left[\begin{array}{ll}
  k_1 & k_2 \\
  -k_2 & k_1
  \end{array}
  \right]  \left[\begin{array}{l}
  x_1 \\
  x_2
  \end{array}
  \right] .
\end{equation}
Thus, the column vector $\textit{{\textbf F}}$ for a s-polarization incidence is
\begin{equation}
\textit{{\textbf F}} = \left[ \begin{array} {c} \widehat{E}_\perp \\ \omega \widehat{H}_\parallel \end{array} \right] .
\label{F}
\end{equation}
Correspondingly, the matrix $M_m$ is
\begin{equation}
  M_m = \left[ \begin{array} {cc}
  0 & \mu_0 \\ \omega^2 \varepsilon_m - \kpara^2 / \mu_0 \quad& 0 \end{array} \right] \, .
\label{Mm}
\end{equation}
For the effective medium in the right panel of Fig. \ref{figure1}, the constitutive relations turn out to be
\begin{equation}
\begin{array}{ll}
\widehat{\!\boldsymbol{D}}(\kpara,x_3) \!\!\!&= \,\varepsilon_{\text{eff}}(\omega,\kpara)
\, \, \widehat{\!\boldsymbol{E}}(\kpara,x_3) \!+\! i  K_{\text{eff}}(\omega,\kpara) J
\, \widehat{\!\boldsymbol{H}}(\kpara,x_3) , \\[2mm]
\!\widehat{\boldsymbol{B}}(\kpara,x_3) \!\!\!&=\mu_{\text{eff}}(\omega,\kpara) \,\widehat{\!\boldsymbol{H}}(\kpara,x_3)
\!+\! i  J K_{\text{eff}}(\omega,\kpara) \widehat{\boldsymbol{E}}(\kpara,x_3) ,
\end{array}
\label{liu1}
\end{equation}
where $\varepsilon_{\rm eff}$ and $\mu_{\rm eff}$ are the effective 
permittivity and permeability tensors,
\begin{equation}
\varepsilon_{\rm eff} =
\left[ \begin{array}{ccc} \vspace*{1mm}
\varepsilon_\parallel & 0 & 0 \\
0 & \varepsilon_\perp & 0 \\
0 & 0 & \varepsilon_3
\end{array}
\right] , \quad
\mu_{\rm eff} =
\left[ \begin{array}{ccc} \vspace*{1mm}
\mu_\parallel & 0 & 0 \\
0 & \mu_\perp & 0 \\
0 & 0 & \mu_3
\end{array}
\right] ;
\label{epmueff}
\end{equation}
Matrix $J$ represents the 90 degrees rotation around the $x_3$-axis, and $K_{\text{eff}}$
is the bianisotropic parameter measuring the magnetoelectric coupling effect:
\begin{equation}
J =\left[ \begin{array}{ccc} \vspace*{1mm}
0 & -1 & 0 \\
1 & 0 & 0 \\
0 & 0 & 1
\end{array}
\right] \, ,\quad 
K_{\text{eff}} =
\left[ \begin{array}{ccc} \vspace*{1mm}
K_\parallel & 0 & 0 \\
0 & K_\perp & 0 \\
0 & 0 & 0
\end{array}
\right] \, .
\label{xieff}
\end{equation}
Similarly, we can obtain the matrix $M_{\rm eff}$ by adopting the same calculation process of $M_m$ in a periodic multilayer
\begin{equation}
M_{\rm eff} =  \left[ \begin{array}{cc} \vspace*{1mm}
-i\omega K_\parallel & \mu_\parallel  \\
\omega^2 \ep_\perp -\kpara^2/\mu_3 & i\omega K_\parallel \end{array} \right] .
\label{Meff}
\end{equation}
The exponential function of matrix $M_m$ defines the transfer matrix for the multilayer 
\begin{equation}
T(\omega,\kpara)=\exp[-i M_m h_m]
\end{equation}
while for the effective medium in the whole complex plane, its transfer matrix is
\begin{equation}
T_\text{eff}(\omega,\kpara)=\exp[-i M_\text{eff} h] \, .
\label{Teff}
\end{equation}
They should satisfy the following relation:
\begin{equation}
\exp[-i M_\text{eff} h] = \exp[-i M_2 h_2] \exp[-i M_1 h_1] \; .
\end{equation}
In order to solve this equation, we need to introduce the Baker-Campbell-Hausdorff (BCH) formula \cite{weiss}, which defines an approximation for the product of two exponential functions with matrix as arguments
\begin{equation}
\exp[A] \exp[B] = \exp ( A + B + \llbracket A, B \rrbracket +\dfrac{1}{3} \llbracket B-A , \llbracket B , A \rrbracket \rrbracket + \cdots )
\label{bch}
\end{equation}
here we denote $A+B$ the zeroth order approximation, the commutator $\llbracket A, B \rrbracket=(A B - B A)/2$ the first order approximation, and so forth.
Hence taking $A=-i M_2 h_2$, $B=-i M_1 h_1$, we obtain
\begin{equation}
\begin{array}{ll}
-i M_\text{eff} h & = -i (M_2 h_2 + M_1 h_1) + \llbracket -i M_2 h_2, -i M_1 h_1 \rrbracket
+\dfrac{1}{3} \llbracket -i M_2 h_2 + i M_1 h_1 , \llbracket -i M_2 h_2 , -i M_1 h_1 \rrbracket \rrbracket + \cdots
\end{array}
\label{approxMeff}
\end{equation}
Assuming a normal incidence whereby $\kpara=\bf{0}$, and taking the zeroth order approximation in (\ref{approxMeff}), we have
\begin{equation}
  M_{\rm eff} \approx M_1 f_1 + M_2 f_2
\label{zero}
\end{equation}
with $f_1=h_1/h, f_2=h_2/h$, which is the result the classical homogenization gives \cite{bensoussan}:
\begin{equation}
  \ep_\perp = \ep_1 f_1 + \ep_2 f_2, \quad  \mu_\parallel =\mu_0, \quad  K_\parallel=0 \, .
\end{equation}
One should note that classical homogenization does not capture any artificial magnetism or bianisotropy effect. 

Next, taking up to the first order approximation in (\ref{approxMeff}) we obtain
\begin{equation}
  M_{\rm eff} \approx M_1 f_1 + M_2 f_2 - i \llbracket M_2 f_2,  M_1 f_1 \rrbracket \, , 
\label{first}
\end{equation}
and the effective parameters turn out to be
\begin{equation}
  \ep_\perp(\hat z)= \ep_1 f_1 + \ep_2 f_2, \quad \mu_\parallel(\hat z) = \mu_0, \quad K_\parallel(\hat z)={\hat z} (\ep_1-\ep_2)f_1 f_2/2\ep_0 \, ,
\end{equation}
which produces a magnetoelectric coupling parameter $K$ depending on the frequency: here, $\hat z = zh \sqrt{\varepsilon_0 \mu_0}$ 
is the normalized complex frequency (the complex frequency is $z = \omega + i \eta$, its real part being the real frequency $\omega$).

The second order approximation in (\ref{approxMeff}) yields
\begin{equation}
\begin{array}{ll}
M_\text{eff}& \approx (M_2 f_2 + M_1 f_1) -i \llbracket M_2 f_2, M_1 f_1 \rrbracket
+\dfrac{1}{3} \llbracket M_2 f_2 - M_1 f_1 , \llbracket -i M_2 f_2 , -i M_1 f_1 \rrbracket \rrbracket  .
\end{array}
\label{second}
\end{equation}
Based on this, we can finally obtain the analytic expressions for effective parameters in equation (11) of the letter, 
where the artificial magnetism and high order corrections to permittivity appear. It should be noted that, under a 
normal incidence, the s-polarization and p-polarization waves are the same under the isotropic system in the plane $(x_1, x_2)$, i.e. $\ep_\parallel(\hat z)=\ep_\perp(\hat z)$, $\mu_\parallel(\hat z)=\mu_\perp(\hat z)$ and $K_\parallel(\hat z)=K_\perp(\hat z)$.

Based on the BCH formula, we can push our algorithm to any order approximation. Here we consider up to the sixth order approximation where the effective parameters are
\begin{align}
{\hat \ep_\parallel(\hat z)}={\hat \ep_\perp(\hat z)} &={\hat \ep_1} f_1+{\hat \ep_2} f_2+\dfrac{{\hat z}^2}{6} f_1 f_2 ({\hat \ep_1}-{\hat \ep_2})({\hat \ep_1} f_1-{\hat \ep_2} f_2) \notag \\
&-\dfrac{{\hat z}^4}{90}f_1 f_2 ({\hat \ep_1}-{\hat \ep_2})\left[({\hat \ep_2} f_2^2-{\hat \ep_1} f_1^2)({\hat \ep_1} f_1+{\hat \ep_2} f_2)+3f_1 f_2 ({\hat \ep_2}^2 f_2-{\hat \ep_1}^2 f_1)\right] \notag\\
&+\dfrac{{\hat z}^6}{3780} f_1 f_2 ({\hat \ep_1} -{\hat \ep_2}) (4 {\hat \ep_1}^3 f_1^5 + 18 {\hat \ep_1}^3 f_1^4 f_2 + 27{\hat \ep_1}^3 f_1^3 f_2^2 + 6{\hat \ep_1}^2 {\hat \ep_2} f_1^4 f_2 + 8{\hat \ep_1}^2 {\hat \ep_2} f_1^3 f_2^2 + 3{\hat \ep_1}^2 {\hat \ep_2} f_1^2 f_2^3 \notag\\[3mm]
&- 3{\hat \ep_1} {\hat \ep_2}^2 f_1^3 f_2^2 - 8{\hat \ep_1} {\hat \ep_2}^2 f_1^2 f_2^3 - 6{\hat \ep_1}{\hat \ep_2}^2 f_1 f_2^4 - 27{\hat \ep_2}^3 f_1^2 f_2^3 - 18{\hat \ep_2}^3 f_1 f_2^4 - 4 {\hat \ep_2}^3 f_2^5) \, , \notag\\
\mu_\parallel(\hat z)=\mu_\perp(\hat z) &= \mu_0-\dfrac{{\hat z}^2}{6} \mu_0 f_1 f_2 ({\hat \ep_1}-{\hat \ep_2})(f_1-f_2) \notag \\
&+\dfrac{{\hat z}^4}{90} \mu_0 f_1 f_2 ({\hat \ep_1}-{\hat \ep_2})\left[({\hat \ep_2}f_2^2-{\hat \ep_1}f_1^2)+3f_1f_2({\hat \ep_1}f_2-{\hat \ep_2}f_1)\right] \notag \\
&-\dfrac{{\hat z}^6}{3780}\mu_0 f_1 f_2 ({\hat \ep_1} -{\hat \ep_2}) (4{\hat \ep_1}^2 f_1^5 + 6 {\hat \ep_1}^2 f_1^4 f_2 - 3{\hat \ep_1}^2 f_1^3 f_2^2 - 27{\hat \ep_1}^2 f_1^2 f_2^3 + 18 {\hat \ep_1} {\hat \ep_2} f_1^4 f_2 + 8{\hat \ep_1} {\hat \ep_2} f_1^3 f_2^2 \notag\\[3mm]
&- 8{\hat \ep_1} {\hat \ep_2} f_1^2 f_2^3 - 18{\hat \ep_1} {\hat \ep_2} f_1 f_2^4 + 27{\hat \ep_2}^2 f_1^3 f_2^2 + 3{\hat \ep_2}^2 f_1^2 f_2^3 - 6{\hat \ep_2}^2 f_1 f_2^4 - 4{\hat \ep_2}^2 f_2^5) \, , \notag \\
K_\parallel(\hat z)=K_\perp(\hat z)&=\dfrac{{\hat z}}{2} ({\hat \ep_1}-{\hat \ep_2})f_1 f_2+\dfrac{{\hat z}^3}{12}({\hat \ep_1}^2-{\hat \ep_2}^2)f_1^2 f_2^2 \notag\\
&-\dfrac{{\hat z}^5}{180}f_1^2 f_2^2({\hat \ep_1} -{\hat \ep_2})({\hat \ep_1}^2 f_1^2 + 3{\hat \ep_1}^2 f_1f_2 + {\hat \ep_1} {\hat \ep_2} f_1^2 + 2{\hat \ep_1} {\hat \ep_2} f_1 f_2 + {\hat \ep_1} {\hat \ep_2} f_2^2 + 3{\hat \ep_2}^2 f_1 f_2 + {\hat \ep_2}^2 f_2^2) \, , 
\label{effpara}
\end{align}
where $\hat \ep_i=\ep_i/\ep_0$ is the relative permittivity ($\varepsilon_0$ is the vacuum permittivity).
It is clear from this expression that all effective parameters depend crucially upon frequency and, permittivity and filling fraction contrasts.

\begin{figure}[!h]
    \centering
    \includegraphics[scale=0.5]{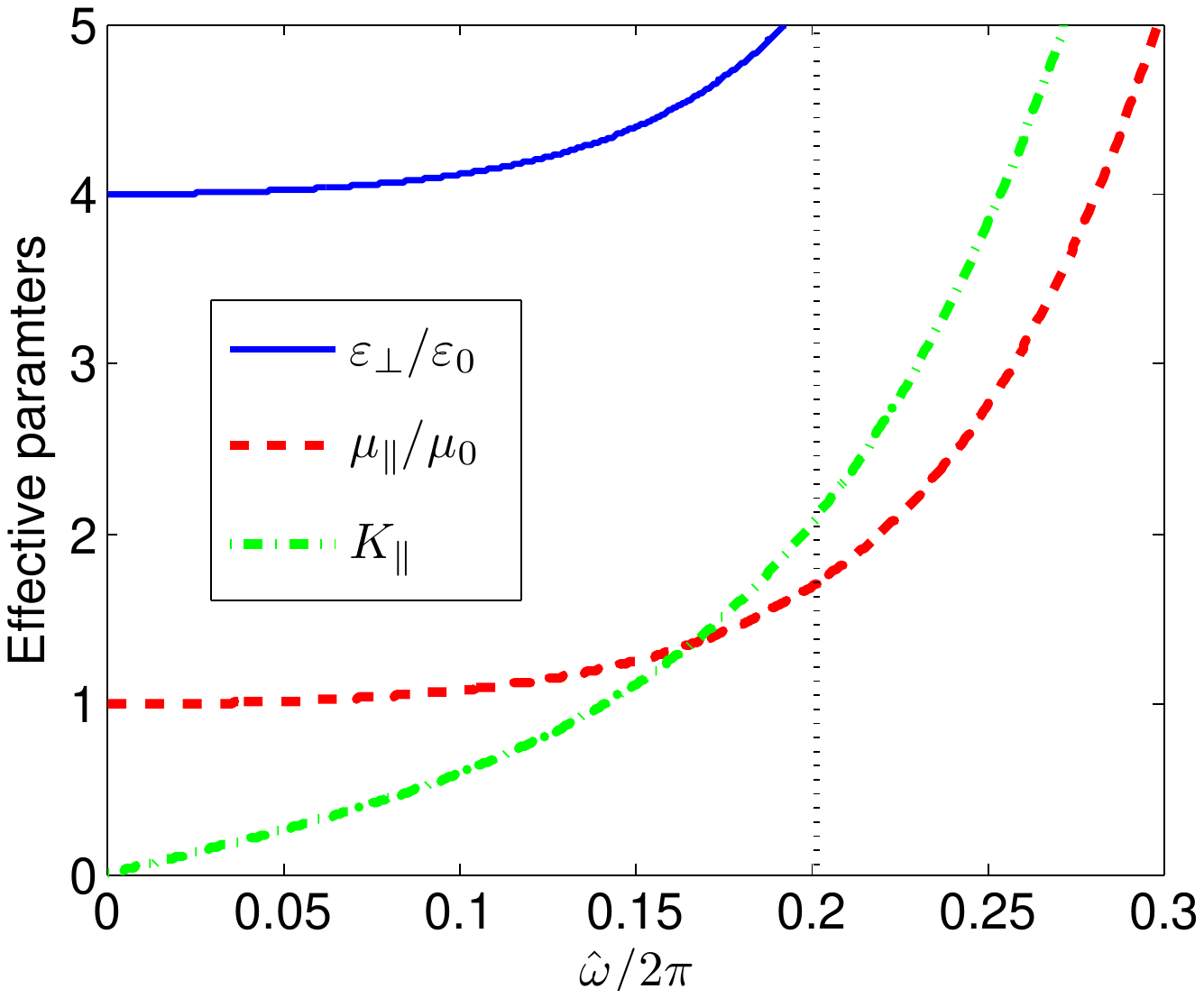}
    \caption{Effective parameters versus normalized real frequency}
    \label{figure2}
\end{figure}

The curves of these effective parameters versus normalized frequency are plotted in Fig. \ref{figure2}, where we have assumed the two dielectrics in the unit cell of the periodic multilayer are Glass and Silicon respectively, with
\begin{equation}
  \hat \ep_1=2,\quad \hat \ep_2=12,\quad f_1=0.8, \quad f_2=0.2 \, .
\end{equation}
It can be observed that all the effective parameters increase along with the frequency, the effective permittivity takes 
the values larger than 1(artificial magnetism), while $K$ describing the magnetoelectric coupling is always non-zero. 
One should note that all correcting terms in ${\hat \ep_\parallel(\hat z)} [={\hat \ep_\perp(\hat z)}]$ and $\mu_\parallel(\hat z) 
[=\mu_\perp(\hat z)]$ are positive.

\subsection{BCH formula for the 3 layers' case}
For a periodic multilayer with a unit cell consisting 3 dielectric layers, the relation between the transfer matrices of multilayers and its effective medium will be
\begin{equation}
\exp[X]=\exp[A]\exp[B]\exp[A] \, ,
\label{bch3}
\end{equation}
which is a product of three exponential functions with matrix argument. In order to derive the expressions for $X$, we can take an iteration of BCH formula of (\ref{bch}) using
\begin{equation}
\exp[C]=\exp[A]\exp[B]
\label{bch3-1}
\end{equation}
with
\begin{equation}
 C = A + B + \dfrac{1}{2} \llbracket A, B\rrbracket+\dfrac{1}{12} \llbracket A, \llbracket A,B \rrbracket \rrbracket-\dfrac{1}{12} \llbracket B, \llbracket A,B \rrbracket \rrbracket  \cdots
\end{equation}
Then equation (\ref{bch3}) becomes
\begin{equation}
\exp[X]=\exp[C]\exp[A] \, ,
\label{bch3-2}
\end{equation}
and hence
\begin{equation}
\begin{array}{l}
  X=A+B+A+\dfrac{1}{2} \llbracket A,B\rrbracket +\dfrac{1}{2} \llbracket A+B,A\rrbracket+\dfrac{1}{12} \llbracket A, \llbracket A,B \rrbracket \rrbracket-\dfrac{1}{12} \llbracket B, \llbracket A,B \rrbracket \rrbracket \\[3mm]
  + \dfrac{1}{4} \llbracket \llbracket A, B\rrbracket,A\rrbracket
  + \dfrac{1}{12} \llbracket A+B, \llbracket A+B, A\rrbracket \rrbracket - \dfrac{1}{12} \llbracket A, \llbracket A+B, A\rrbracket \rrbracket+\cdots
\end{array}
\label{bch3-3}
\end{equation}
Similarly, we take up to the fourth order approximation under a normal incident in s-polarization, the effective parameters are
\begin{align}
{\hat \ep_\parallel}(\hat z)={\hat\ep_\perp}(\hat z) &= 2{\hat\ep_1} f_1+{\hat\ep_2} f_2 - \dfrac{{\hat z}^2}{3}f_1f_2 ({\hat\ep_1}-{\hat\ep_2})({\hat\ep_1} f_1+{\hat\ep_2} f_2) \notag \\
&-\dfrac{{\hat z}^4}{45}f_1 f_2({\hat\ep_1}-{\hat\ep_2})(7{\hat\ep_1}^2 f_1^3 + 3 {\hat\ep_1}^2 f_1^2 f_2 + 11{\hat\ep_1} {\hat\ep_2} f_1^2 f_2 + 2{\hat\ep_1} {\hat\ep_2} f_1 f_2^2 + {\hat\ep_2}^2 f_2^3+ 6 {\hat\ep_2}^2 f_1 f_2^2) \, , \notag \\
\mu_\parallel(\hat z)=\mu_\perp(\hat z)&= \mu_0 + \dfrac{{\hat z}^2}{3}\mu_0f_1f_2 ({\hat \ep_1}-{\hat \ep_2})(f_1+f_2) \, , \notag \\
&+\dfrac{{\hat z}^4}{45}\mu_0f_1 f_2 ({\hat \ep_1}-{\hat \ep_2})(7 {\hat \ep_1} f_1^3 + 11{\hat \ep_1} f_1^2f_2 + 6{\hat \ep_1}f_1 f_2^2+ 3 {\hat \ep_2}f_1^2f_2 + {\hat \ep_2}f_2^3 + 2{\hat \ep_2}f_1f_2^2) \notag \\
K_\parallel(\hat z)=K_\perp(\hat z)&=0 \, .
\label{3lay2}
\end{align}
The results indicate that the magnetodielectric coupling vanishes in the three layers' case, which can be explained by the fact that: the odd 
order approximations in (\ref{bch3-3}) are all equal to zero due to the anti-commutation in the commutators 
($\llbracket A,B\rrbracket=-\llbracket B,A\rrbracket$).

\subsection{Divergence of the power expansion at the band gap edge}

In this section, we show that expressions of the effective parameters introduced by the power expansion 
in previous sections are no longer valid at the first band gap edge. 

First, we expand the transfer matrix in (\ref{Teff}) with Taylor series:
\begin{equation}
T_{\rm eff} = \exp(-iM_{\rm eff}h) 
=\sum_{p=0}^{\infty} (-i)^{2p} \dfrac{M_{\rm eff}^{2p}h^{2p}}{(2p)!}+\sum_{p=0}^{\infty} (-i)^{2p+1} \dfrac{M_{\rm eff}^{2p+1}h^{2p+1}}{(2p+1)!} \, .
\label{T}
\end{equation}
According to (\ref{Meff}) where the real frequency $\omega$ is replaced by the complex one $z$, we have
\begin{equation}
M_{\rm eff}^2 =
 \left[ \begin{array}{cc} \vspace*{1mm}
 -iz K_\parallel/c_0 & \mu_\parallel  \\
z^2 \ep_\perp -\kpara^2/\mu_3 & iz K_\parallel/c_0
 \end{array} \right]^2
=\left( z^2\ep_\perp \mu_\parallel-\dfrac{z^2}{c_0^2}K_\parallel^2-\dfrac{\mu_\parallel}{\mu_3} \kpara^2 \right)
  \left[ \begin{array}{cc} \vspace*{1mm}
 1 & 0\\
 0 & 1\\
 \end{array} \right] = k_{\rm eff}^2 \, , 
\label{Mte2}
\end{equation}
with 
\begin{equation}
k_{\rm eff}^2=z^2\ep_\perp \mu_\parallel-\dfrac{z^2}{c_0^2}K_\parallel^2-\dfrac{\mu_\parallel}{\mu_3} \kpara^2 \, .
\label{n2}
\end{equation}
Let us plug (\ref{Mte2}) into (\ref{T})
and, considering Taylor series of the $\sin-\cos$ functions
\begin{equation}
    \sin(A)=\sum_{n=0}^{\infty} \dfrac{(-1)^n}{(2n+1)!} A^{2n+1} \, , \quad 
    \cos(A)=\sum_{n=0}^{\infty} \dfrac{(-1)^n}{(2n)!} A^{2n} \, ,
\label{sincos}
\end{equation}
we obtain
\begin{equation}
T_{\rm eff}
=\cos(k_{\rm eff} h)
-i\dfrac{M_{\rm eff}}{k_{\rm eff}} \sin(k_{\rm eff} h)
=\left[ \begin{array}{cc} \vspace*{1mm}
 \cos(k_{\rm eff} h)-\dfrac{z K_\parallel}{c_0} \dfrac{\sin(k_{\rm eff} h)}{k_{\rm eff}} & 
-i\mu_\parallel \dfrac{\sin(k_{\rm eff} h)}{k_{\rm eff}}\\
 -i(z^2\varepsilon_\perp-\dfrac{\kpara^2}{\mu_3} )\dfrac{\sin(k_{\rm eff} h)}{k_{\rm eff}} & 
\cos(k_{\rm eff} h)+\dfrac{z K_\parallel}{c_0} \dfrac{\sin(k_{\rm eff}}{k_{\rm eff}}\\
 \end{array} \right] \, .
\label{Tte}
\end{equation}
Let us now write the characteristic matrix \cite{optics} of the dielectrics of the periodic multilayer as follows 
\begin{equation}
T_i=\left[\begin{array}{cc}
\cos{(k_ih_i)} & \dfrac{-i\zeta_i}{z}\sin{(k_ih_i)}\\
-i\dfrac{z}{\zeta_i} \sin{(k_ih_i)} & \cos{(k_ih_i)}
\end{array}\right], \quad k_i^2=z^2\ep_i \mu_i-\kpara^2, \quad \zeta_i^2=\dfrac{\mu_i}{\ep_i} \, .
\label{tp}
\end{equation}
The transfer matrix of the unit cell consisting of three dielectric layers ($\ep_1=\ep_3$, $h_1=h_3$) is
\begin{equation}
  T=T_1 T_2 T_1 \, .
\end{equation}
Since the dispersion law is defined by the trace of transfer matrix $T$, we have
\begin{equation}
  \tr(T)/2 =\left\{
  \begin{array}{ll}
  \cos(k_{\rm eff} h) &\qquad {\rm effective \, medium,} \\
  \cos{(2k_1h_1)}\cos{(k_2h_2)}-\dfrac{1}{2}(\dfrac{\zeta_2}{\zeta_1}+\dfrac{\zeta_1}{\zeta_2})\sin{(2k_1h_1)}\sin{(k_2h_2)} & \qquad {\rm multilayer\: stack\:with\:three\: layers.}
  \end{array} \right.
\label{trT}
\end{equation}
The dispersion law of the effective medium is depicted in dashed line [here we have taken up to the 20th order approximation in (\ref{bch3-3}), 
and substitute the analytic expressions of the effective parameters into (\ref{n2}) and then to (\ref{trT})], as well as that of the multilayers 
in solid line in Fig. \ref{figure3}, assuming a normal incidence within s-polarization. One can see that the dashed line diverges 
when frequency tends to the lower edge of the first stop band, Sdenoted by $\hat \omega_1$. In other words, the expressions of these 
effective parameters are not valid any more further the first band gap in multilayers.
\begin{figure}[!h]
    \centering
    \includegraphics[scale=0.55]{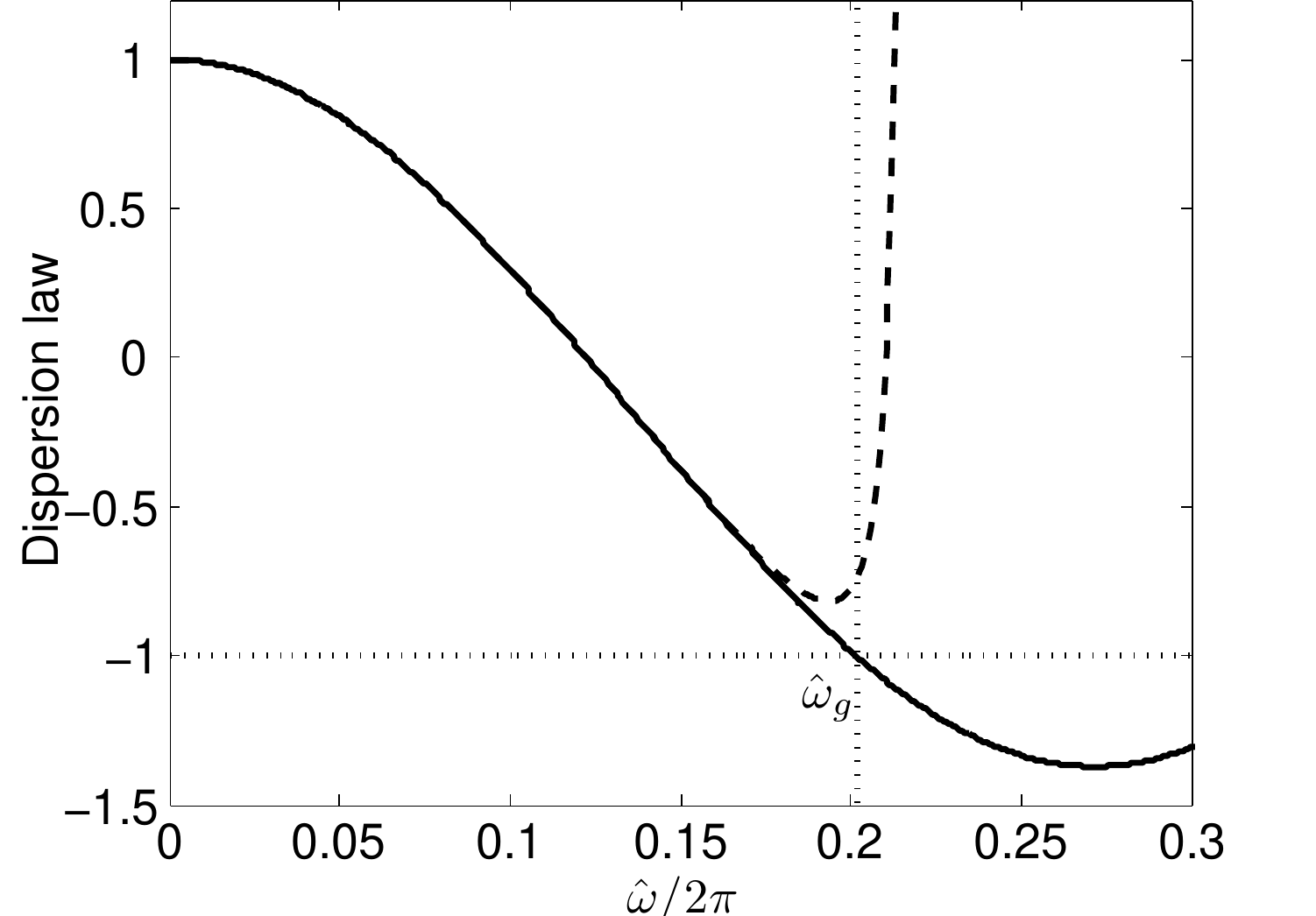}
    \caption{Dispersion laws of multilayers (solid line) and effective medium (dashed line in 20th order approximation).}
    \label{figure3}
\end{figure}

\subsection{Proof of iii) of the main result: Effective index and surface impedance at infinite frequency}

As it has been shown in the previous section, the effective permittivity, permeability and magnetoelectric parameters are no longer valid when $\omega>\omega_1$. To overcome this pitfall, we consider the set of effective parameters made of propagation index $n_{\rm eff}$ 
and surface impedance $\zeta_{\rm eff}$. 

We consider a periodic multilayer with a unit cell consisting of three layers, and comparing $T=T_1T_2T_1$ with the transfer matrix of the effective medium in (\ref{Tte}), 
we find
\begin{equation}
\begin{array}{ll}
  {\rm a}=\cos{[zn_{\rm eff}h]}&=\cos{(2k_1h_1)}\cos{(k_2h_2)}-\dfrac{1}{2}(\dfrac{\zeta_2}{\zeta_1}+\dfrac{\zeta_1}{\zeta_2})\sin{(2k_1h_1)}\sin{(k_2h_2)} \\
  &=\cos{(2k_1h_1+k_2h_2)}-\dfrac{1}{2}(\dfrac{\zeta_2}{\zeta_1}+\dfrac{\zeta_1}{\zeta_2}-2)\sin{(2k_1h_1)}\sin{(k_2h_2)} \, ,
  \label{index}
\end{array}
\end{equation}
and
\begin{equation}
\begin{array}{l}
  z  b =-i 2\zeta_1 \sin{(k_1h_1)}\cos{(k_1h_1)}\cos{(k_2h_2)}+i\dfrac{\zeta_1^2}{\zeta_2}\sin{(k_1h_1)}^2\sin{(k_2h_2)}-i\zeta_2\cos{(\beta_1)}^2\sin{(k_2h_2)} \, , \\[2mm]
  d/z=-i\dfrac{2}{\zeta_1}\sin{(k_1h_1)}\cos{(k_1h_1)}\cos{(k_2h_2)}+i\dfrac{\zeta_2}{\zeta_1^2}\sin{(k_1h_1)}^2\sin{(k_2h_2)}-i\dfrac{1}{\zeta_2}\cos{(k_1h_1)}^2\sin{(k_2h_2)} \, .
\end{array}
\end{equation}
When frequency $z=\omega+i\eta=|z|e^{i\phi}$ tends to infinity by $|z|\rightarrow\infty$, Euler's formulae
\begin{equation}
  \sin{z}=\dfrac{e^{iz}-e^{-iz}}{2i} \, ,\quad
  \cos{z}=\dfrac{e^{iz}+e^{-iz}}{2} \, , 
\end{equation}
simplify into
\begin{equation}
  \sin{z}\sim-\dfrac{e^{-iz}}{2i} \, ,\quad
  \cos{z}\sim\dfrac{e^{-iz}}{2} \, ,
\label{euler}
\end{equation}
since $e^{iz}\rightarrow 0$.
Equation (\ref{index}) becomes
\begin{equation}
  \exp{(-izn_{\rm eff}h)}\sim\exp{(-i2k_1h_1-ik_2h_2)}+A \exp{(-i2k_1h_1-ik_2h_2)}
\end{equation}
with $A=(\zeta_2/\zeta_1+\zeta_1/\zeta_2-2)/4$ a constant. Furthermore,
\begin{equation}
  izn_{\rm eff}h\sim i(2k_1h_1+k_2h_2)-\ln{(1+A)}
\end{equation}
when $|z|\rightarrow \infty$, $k_m \rightarrow z n_m$, then
\begin{equation}
  n_{\rm eff} \sim (2z n_1h_1+zn_2h_2)/(z h) + \dfrac{i}{zh} \ln[1 + A] \underset{\infty}{\rightarrow} \langle n \rangle \, .
\end{equation}
This proves that the limit of the refractive index at infinite frequency is equal to the mean of indices of a 
unit cell, as stated in the letter.

The surface impedance is defined by
\begin{align}
  \zeta_{\rm eff}^2=z^2\dfrac{{\rm b}}{{\rm d}}&=\dfrac{-i 2\zeta_1 \sin{(k_1h_1)}\cos{(k_1h_1)}\cos{(k_2h_2)}+i\dfrac{\zeta_1^2}{\zeta_2}\sin{(k_1h_1)}^2\sin{(k_2h_2)}-i\zeta_2\cos{(\beta_1)}^2\sin{(k_2h_2)}}
  {-i\dfrac{2}{\zeta_1}\sin{(k_1h_1)}\cos{(k_1h_1)}\cos{(k_2h_2)}+i\dfrac{\zeta_2}{\zeta_1^2}\sin{(k_1h_1)}^2\sin{(k_2h_2)}-i\dfrac{1}{\zeta_2}\cos{(k_1h_1)}^2\sin{(k_2h_2)}} \nonumber\\[2mm]
  &=\zeta_1^2 \dfrac{\sin{(k_2h_2)}\left[{\zeta_1^2}\sin^2{(k_1h_1)}-{\zeta_2^2}\cos^2{(k_1h_1)}\right]-\zeta_1\zeta_2\sin{(2k_1h_1)}\cos{(k_2h_2)}}
  {\sin{(k_2h_2)}\left[{\zeta_2^2}\sin^2{(k_1h_1)}-{\zeta_1^2}\cos^2{(k_1h_1)}\right]-{\zeta_1\zeta_2}\sin{(2k_1h_1)}\cos{(k_2h_2)}} .
\label{imp}
\end{align}
Plugging (\ref{euler}) into (\ref{imp}),
\begin{align}
  \zeta_{\infty}&=\lim\limits_{|z| \to \infty }{\zeta_{\rm eff}} =\lim\limits_{|z| \to \infty }{z\sqrt{\dfrac{{\rm b}}{{\rm d}}}}  \notag \\
  &=\lim\limits_{|z| \to \infty }{\left\{{\zeta_1^2} \dfrac{\dfrac{e^{ik_2h_2}}{2i}\left[\zeta_1^2\dfrac{-e^{2ik_1h_1}}{4}-\zeta_2^2\dfrac{e^{2ik_1h_1}}{4}\right]-\zeta_1 \zeta_2\dfrac{e^{2ik_1h_1}}{2i}\dfrac{e^{ik_2h_2}}{2}}
  {\dfrac{e^{ik_2h_2}}{2i}\left[\zeta_2^2\dfrac{-e^{2ik_1h_1}}{4}-\zeta_1^2\dfrac{e^{2ik_1h_1}}{4}\right]-\zeta_1 \zeta_2\dfrac{e^{2ik_1h_1}}{2i}\dfrac{e^{ik_2h_2}}{2}}\right\}}^{1/2}
  =\zeta_1 \, , 
\label{imp2}
\end{align}
the surface impedance is shown to converge at infinity to the impedance of layer $1$.

More generally, for a unit cell consisting of $m$ layers with permittivity $\varepsilon_m$ and thickness $h_m$, the transfer matrix is
\begin{equation}
  T=\prod\limits_{i=1}^{m}{T_i} \, .
  \label{tu}
\end{equation}
When $|z| \rightarrow \infty$, (\ref{tp}) becomes
\begin{equation}
   T_i \sim \dfrac{\exp{(-ik_ih_i)}}{2} \left[\begin{array}{cc}
    1 & -i\zeta'_i \\
    -i\dfrac{1}{\zeta'_i} & 1
  \end{array}\right]
\end{equation}
with $\zeta'_i={\zeta_i/z}$ ($i = 1, \dots m$). According to (\ref{tu}), we have
\begin{equation}
  T_{11}\sim\exp{[-i z (k_1 h_1 + k_2 h_2 \dots + k_m h_m)]} (1+A)
\end{equation}
where $A$ is an algebraic function of ratios $\zeta_i/\zeta_j = \zeta'_i/\zeta'_j$ ($i,j = 1, \dots m$). Finally, we obtain
\begin{equation}
  \exp{(-izn_{\rm eff}h)}\sim\exp{[-i z (k_1 h_1 + k_2 h_2 \dots + k_m h_m)]}(1+A) \, , 
\end{equation}
and
\begin{equation}
  n_{\rm eff}(z) \sim (k_1 h_1 + k_2 h_2 \dots + k_m h_m)/(z h) + \dfrac{i}{zh} \ln[1 + A] \underset{\infty}{\rightarrow} \langle n \rangle \, .
\end{equation}

The same limit for the effective parameters $n_{\text{eff}}(z)$ and $\zeta_{\text{eff}}(z)$ can be obtain 
when Re$(z) = \omega \longrightarrow \infty$. In this case, adding an arbitrary small imaginary part to 
dielectric constants $\varepsilon_m$, the above derivation applies \textit{mutatis mutandis}. 

\subsection{Kramers and Kronig formula for the effective surface impedance}

From the definition of effective surface impedance given by equation (14) in the letter, we know that the function
\begin{equation}
F(z,\kpara)=i [ \zeta_{\rm eff}(z,\kpara)-\zeta_{\infty}]
\label{imp2}
\end{equation}
is analytic in the upper half-plane, and tends to $0$ when $|z|\rightarrow \infty$. Applying the Cauchy integral formula to function $F(z)$, a relationship 
equivalent to Kramers and Kronig formula can be obtained:
\begin{equation}
  \left[i(\zeta_{\rm eff}(z,\kpara)-\zeta_{\infty})\right]=\dfrac{1}{\pi}\int\limits_{-\infty}^{\infty} d\nu \dfrac{{\rm Im} \left[i(\zeta_{\rm eff}(\nu,\kpara)-\zeta_{\infty})\right]}{\nu-z}
\label{Cf}
\end{equation}
The real and imaginary parts of $F(z)$ are shown in Fig. \ref{figure4} in solid lines, while the line denoted by plus markers is obtained from (\ref{Cf}), while a small dissipation has been introduced in the permittivities of the dielectrics to enforce the convergence of the effective surface impedance. The consistency between those two curves of the real part of $F(z)$ confirms that $\zeta_{\rm eff}$ must satisfy the Kramers and Kronig formula.
\begin{figure}[!h]
    \centering
    \includegraphics[scale=0.6]{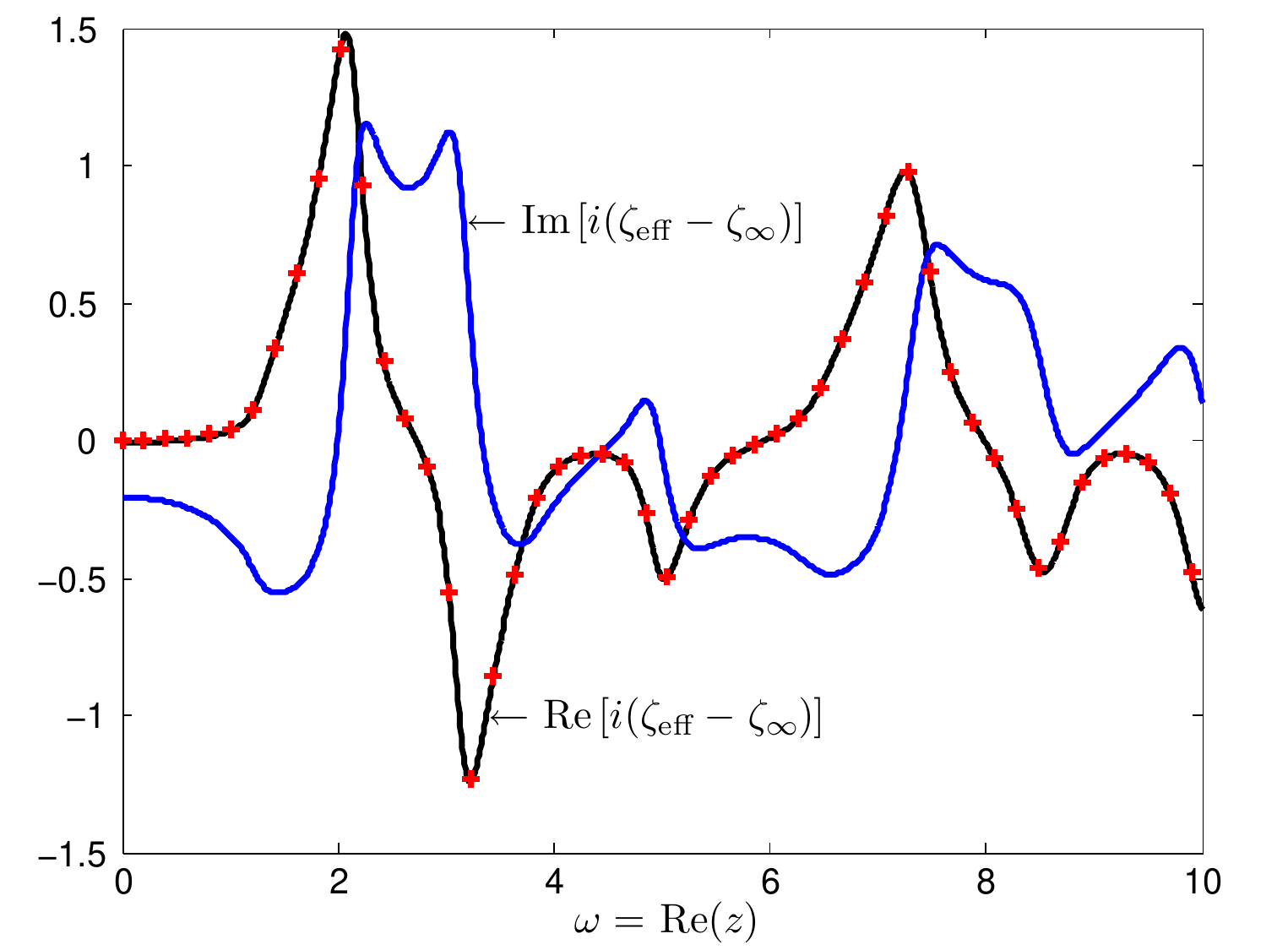}
    \caption{Real and imaginary parts (solid lines) of effective surface impedance $i(\zeta_{\rm eff}-\zeta_{\infty})$ for $z=\omega+0.1*i$, deduced from (\ref{Cf}); while Cauchy integral formula unveils ${\rm Re} \left[F(z)\right]$ in plus markers; here $\ep_1=\ep_3=(2+0.1*i)\varepsilon_0$, $\ep_2=(12+0.1*i)\varepsilon_0$, $f_1=f_3=0.4$, $f_2=0.2$.}
    \label{figure4}
\end{figure}

}

\end{document}